# On X-ray scattering model for single particles Part II: Beyond protein crystallography

Aliakbar Jafarpour

*Dept. of Biomolecular Mechanisms, Max-Planck Inst. for Medical Research, Jahnstr 29, 69120 Hiedelberg, Germany*
*jafarpour.a.j@ieee.org*

**Abstract:** Emerging coherent X-ray scattering patterns of single-particles have shown dominant *morphological* signatures in agreement with predictions of the scattering model used for conventional protein crystallography. The key question is if and to what extent these scattering patterns contain *volumetric* information, and what model can retrieve it. This contribution is *Part 2* out of two reports, in which we seek to clarify the assumptions of some different regimes and models of X-ray scattering and their implications for single particle imaging. In *Part 1*, basic concepts and existing scattering models along with their implications for nanocrystals, and also the misconception of using Diffraction Theory for volumetric scattering were addressed. Here in *Part 2*, specific challenges ahead of single particle imaging are addressed. Limitations of the conventional scattering model in the test case of a sphere, schemes for improving this common scattering model, ambiguities in meaning and properties of "density map" and non-classical effects, the crucial role of electromagnetic boundary conditions, the uniqueness of X-ray scattering inverse problems, and additional vulnerabilities of phase retrieval and its relation with "resolution" are discussed. We raise concerns about the *unverified* use of the common scattering model of protein crystallography for arbitrary objects, and also leaving fundamental questions such as meaning/uniqueness/properties of the sought 3D profiles to phase retrieval algorithms.

**References and Links**

1. J. Miao, P. Charalambous, J. Kirz, and D. Sayre, "Extending the methodology of X-ray crystallography to allow imaging of micrometre-sized non-crystalline specimens," Nature **400**, 342 (1999).
2. S. Kassemeyer, A. Jafarpour, L. Lomb, J. Steinbrener, A.V. Martin, and I. Schlichting "Optimal mapping of x-ray laser diffraction patterns into three dimensions using routing algorithms," Phys. Rev. E **88**, 042710 (2013).
3. M. M. Seibert, T. Ekeberg, F. R. N. C. Maia, M. Svenda, J. Andreasson, O. Jönsson, D. Odic, B. Iwan, A. Rocker, D. Westphal, M. Hantke, D. P. DePonte, A. Barty, J. Schulz, L. Gumprecht, N. Coppola, A. Aquila, M. Liang, T. A. White, A. Martin, C. Caleman, S. Stern, C. Abergel, V. Seltzer, J. Claverie, C. Bostedt, J. D. Bozek, S. Boutet, A. A. Miahnahri, M. Messerschmidt, J. Krzywinski, G. Williams, K. O. Hodgson, M. J. Bogan, C. Y. Hampton, R. G. Sierra, D. Starodub, I. Andersson, S. Bajt, M. Barthelmess, J. C. H. Spence, P. Fromme, U. Weierstall, R. Kirian, M. Hunter, R. B. Doak, S. Marchesini, S. P. Hau-Riege, M. Frank, R. L. Shoeman, L. Lomb, S. W. Epp, R. Hartmann, D. Rolles, A. Rudenko, C. Schmidt, L. Foucar, N. Kimmel, P. Holl, B. Rudek, B. Erk, A. Hömke, C. Reich, D. Pietschner, G. Weidenspointner, L. Strüder, G. Hauser, H. Gorke, J. Ullrich, I. Schlichting, S. Herrmann, G. Schaller, F. Schopper, H. Soltau, K. Kühnel, R. Andritschke, C. Schröter, F. Krasniqi, M. Bott, S. Schorb, D. Rupp, M. Adolph, T. Gorkhover, H. Hirsemann, G. Potdevin, H. Graafsma, B. Nilsson, H. N. Chapman, and J. Hajdu, "Single mimivirus particles intercepted and imaged with an X-ray laser," Nature **470**, 78 (2011).
4. A. Authier, "Dynamical theory of X-ray diffraction," International Tables for Crystallography **B**, 534 (2006).
5. G. Thorkildsen and H. B. Larsen, "X-ray diffraction in perfect t × l crystals. Rocking curves," Acta Cryst **A55**, 840 (1999).
6. G. Gouesbeta, J. A. Lockb, and G. Gréhana, "Generalized Lorenz–Mie theories and description of electromagnetic arbitrary shaped beams: Localized approximations and localized beam models, a review," J. Quant. Spectr. Rad. Transfer **112**, 1 (2011).
7. D. Rupp (TU Berlin, 2013), http://opus4.kobv.de/opus4-tuberlin/frontdoor/index/index/docId/3761

8. C. Bostedt, E. Eremina, D. Rupp, M. Adolph, H. Thomas, M. Hoener, A. R. B. de Castro, J. Tiggesbäumker, K.-H. Meiwes-Broer, T. Laarmann, H. Wabnitz, E. Plönjes, R. Treusch, J. R. Schneider, and T. Möller, "Ultrafast X-Ray Scattering of Xenon Nanoparticles: Imaging Transient States of Matter," Phys. Rev. Lett. **108**, 093401 (2012).
9. L. Méès, G. Gouesbet, and G. Gréhan, "Interaction between femtosecond pulses and a spherical microcavity: internal fields," Opt. Commun. **199**, 33 (2001).
10. L. Méès, G. Gouesbet, and G. Gréhan, "Transient internal and scattered fields from a multi-layered sphere illuminated by a pulsed laser," Opt. Commun. **282**, 4189 (2009).
11. C. F. Bohren and D. R. Huffman, *Absorption and Scattering of Light by Small Particles* (Wiley-VCH, 1998).
12. I. Robinson, F. Pfeiffer, I. Vartanyants, Y. Sun, and Y. Xia, "Enhancement of coherent X-ray diffraction from nanocrystals by introduction of X-ray optics," Opt. Express **11**, 2329 (2003).
13. G. J. Williams, H. M. Quiney, B. B. Dhal, C. Q. Tran, K. A. Nugent, A. G. Peele, D. Paterson, and M. D. de Jonge, "Fresnel Coherent Diffractive Imaging," Phys. Rev. Lett. **97**, 025506 (2006).
14. D. J. Vine, G. J. Williams, B. Abbey, M. A. Pfeifer, J. N. Clark, M. D. de Jonge, I. McNulty, A. G. Peele, and K. A. Nugent, "Ptychographic Fresnel coherent diffractive imaging,". Phys. Rev. A **80**, 063823 (2009).
15. X. Li, C.-Y. Shew, L. He, F. Meilleur, D. A. A. Myles, E. Liu, Y. Zhang, G. S. Smith, K. W. Herwig, R. Pynn, and W.-R. Chen, "Scattering functions of Platonic solids," J. Appl. Cryst. **44**, 545 (2011).
16. H. Kogelnik, *Theory of Optical Waveguides* (Springer Verlag, 1988).
17. C. Xiao and M. G. Rossmann, "Structures of giant icosahedral eukaryotic dsDNA viruses," Current Opinion in Virology **1**, 101 (2011).
18. P. Thibault, V. Elser, C. Jacobsen, D. Shapiro, and D. Sayre, "Reconstruction of a yeast cell from X-ray diffraction data," Acta Cryst. **A62**, 248 (2006).
19. Johnson, "Notes on Green's functions in inhomogeneous media" (MIT, 2011), *http://math.mit.edu/~stevenj/18.303/inhomog-notes.pdf*
20. S. M. Rao, D. Wilton, A. W. Glisson, "Electromagnetic scattering by surfaces of arbitrary shape," IEEE Trans. Antenna Propagat. **30**, 409 (1982).
21. H. Reid (MIT, 2012), http://math.mit.edu/~stevenj/18.369/BEMTalk.20120318.pdf.
22. Y. Han, Z. Cui, and W. Zhao, "Scattering of Gaussian beam by arbitrarily shaped particles with multiple internal inclusions," Opt. Express **20**, 718 (2012).
23. B. Momeni, M. Badieirostami, and A. Adibi, "Accurate and efficient techniques for the analysis of reflection at the interfaces of three-dimensional photonic crystals," JOSA B **24**, 2957 (2007).
24. M. P. Seah, S. J. Spencer, F. Bensebaa, I. Vickridge, H. Danzebrink, M. Krumrey, T. Gross, W. Oesterle, E. Wendler, B. Rheinländer, Y. Azuma, I. Kojima, N. Suzuki, M. Suzuki, S. Tanuma, D. W. Moon, H. J. Lee, H. M. Cho, H. Y. Chen, A. T. S. Wee, T. Osipowicz, J. S. Pan, W. A. Jordaan, R. Hauert, U. Klotz, C. van der Marel, M. Verheijen, Y. Tamminga, C. Jeynes, P. Bailey, S. Biswas, U. Falke, N. V. Nguyen, D. Chandler-Horowitz, J. R. Ehrstein, D. Muller, and J. A. Dura, "Critical review of the current status of thickness measurements for ultrathin $SiO_2$ on Si Part V: Results of a CCQM pilot study," Surface and Interface Analysis **36**, 1269 (2004).
25. R. W. Ziolkowski, J. M. Arnold, and D. M. Gogny, "Ultrafast pulse interactions with two-level atoms," Phys. Rev. A, **52**, 3082 (1995).
26. K. J. Hebert, S. Zafar, E. A. Irene, R. Kuehn, T. E. McCarthy, and E. K. Demirlioglu, "Measurement of the refractive index of thin SiO2 films using tunneling current oscillations and ellipsometry," Appl. Phys. Lett. **68**, 266 (1996).
27. R-J Zhang, Y-M Chen, W-J Lu, Q-Y Cai, Y-X Zheng, and L-Y Chen, "Influence of nanocrystal size on dielectric functions of Si nanocrystals embedded in $SiO_2$ matrix," Appl. Phys. Lett. **95**, 161109 (2009).
28. M. Quniten, "Limitations of Mie's Theory – Size and Quantum Size Effects in Very Small Nanoparticles" in *Optical Properties of Nanoparticle Systems: Mie and beyond* (Wiley VCH, 2011).
29. T. van Buuren, L. N. Dinh, L. L. Chase, W. J. Siekhaus, and L. J. Terminello, "Changes in the Electronic Properties of Si Nanocrystals as a Function of Particle Size," Phys. Rev. Lett. **80**, 3803 (1998).
30. A. Zimina, S. Eisebitt, W. Eberhardt, J. Heitmann, and M. Zacharias, "Electronic structure and chemical environment of silicon nanoclusters embedded in a silicon dioxide matrix," Appl. Phys. Lett. **88**, 163103 (2006).
31. L. Šiller, S. Krishnamurthy, L. Kjeldgaard, B. R. Horrocks, Y. Chao, A. Houlton, A. K. Chakraborty, and M. R. C. Hunt, "Core and valence exciton formation in x-ray absorption, x-ray emission and x-ray excited optical luminescence from passivated Si nanocrystals at the Si $L_{2,3}$ edge," J. Phys.: Condens. Matter **21**, 095005 (2009).
32. L. Landt, K. Klünder, J. E. Dahl, R. M. K. Carlson, T. Möller, and C. Bostedt, "Optical Response of Diamond Nanocrystals as a Function of Particle Size, Shape, and Symmetry," Phys. Rev. Lett. **103**, 047402 (2009).
33. N. Majer, K. Ludge, and E. Scholl, "Maxwell-Bloch approach to four-wave mixing in quantum dot semiconductor optical amplifiers," Numerical Simulation of Optoelectronic Devices (NUSOD), 11th International Conference on, 153 (2011).
34. M. Kolarczik, N. Owschimikow, J. Korn, B. Lingnau,Y. Kaptan, D. Bimberg, E. Schöll, K. Lüdge, and U. Woggon, "Quantum coherence induces pulse shape modification in a semiconductor optical amplifier at room temperature," Nature Communications **4**, 2953 (2013).
35. N. Rohringer, D. Ryan, R. A. London, M. Purvis, F. Albert, J. Dunn, J. D. Bozek, C. Bostedt, A. Graf, R. Hill, S. P. Hau-Riege, and J. J. Rocca, "Atomic inner-shell X-ray laser at 1.46 nanometres pumped by an X-ray free-electron laser," Nature **481**, 10721 (2012).

36. I. R. Al'miev, O. Larroche, D. Benredjem, J. Dubau, S. Kazamias, C. Möller, and A. Klisnick, "Dynamical Description of Transient X-Ray Lasers Seeded with High-Order Harmonic Radiation through Maxwell-Bloch Numerical Simulations," Phys. Rev. Lett. **99**, 123902 (2007).
37. F. Tissandier, S. Sebban, J. Gautier, Ph. Zeitoun, E. Oliva, A. Rousse, and G. Maynard, "Three-dimensional Maxwell-Bloch calculation of the temporal profile of a seeded soft x-ray laser pulse," Appl. Phys. Lett. **101**, 251112 (2012).
38. A. Taflove and S. C. Hangess, *Computational Electrodynamics* (Artech House, 2000).
39. B. Rupp (2009), http://www.ruppweb.org/Xray/comp/scatfac.htm.
40. P. P. Ewald, "Zur Begründung der Kristalloptik, Teil III: Die Kristalloptik der Röntgenstrahlen," Annalen der Physik **359**, 557 (1917).
41. C. M. Reinke, A. Jafarpour, B. Momeni, M. Soltani, S. Khorasani, A. Adibi, Y. Xu, and R. K. Lee, "Nonlinear finite-difference time-domain method for the simulation of anisotropic, $\chi^{(2)}$, and $\chi^{(3)}$ optical effects," Journal of Lightwave Technology **24**, 624 (2006).
42. L. Cheng and D. Y. Kim, "Differential imaging in coherent anti-Stokes Raman scattering microscopy with Laguerre- Gaussian excitation beams," Opt. Express **15**(16), 10123 (2007).
43. B. T. Draine and P. J. Flatau, "Discrete dipole approximation for scattering calculations," J. Opt. Soc. Am. **A11**, 1491 (1994).
44. M. A. Yurkina and A. G. Hoekstrac, "The discrete-dipole-approximation code ADDA: Capabilities and known limitations," J. Quant. Spectrosc. Radiat. Transfer **112**, 2234 (2011).
45. M. A. Yurkina and M. Kahnert, "Light scattering by a cube: Accuracy limits of the discrete dipole approximation and the T-matrix method," J. Quant. Spectrosc. Radiat. Transfer **123**, 176 (2013).
46. B. Rupp, *Biomolecular Crystallography* (Garland Science, 2010).
47. J. Drenth, "Basic diffraction physics," International Tables for Crystallography **F**, 52 (2006).
48. H. Liu, B. K. Poon, D. K. Saldin, J. C. H. Spence, and P. H. Zwart, "Three-dimensional single-particle imaging using angular correlations from X-ray laser data," Acta Cryst. **A69**, 365 (2013).
49. J. C. H. Spence, U. Weierstall, and H. N. Chapman, "X-ray lasers for structural and dynamic biology," Rep. Prog. Phys. **75**, 102601 (2012).
50. X. Yan, N. H. Olson, J. L. Van Etten, M. Bergoin, M. G. Rossmann, and T. S. Baker, "Structure and assembly of large lipid-containing dsDNA viruses," Nature Structural Biology **7**, 101 (2000).
51. M. V. Cherrier, V. A. Kostyuchenko, C. Xiao, V. D. Bowman, A. J. Battisti, X. Yan, P. R. Chipman, T. S. Baker, J. L. Van Etten, and M. G. Rossmann, "An icosahedral algal virus has a complex unique vertex decorated by a spike," PNAS **106**, 11085 (2009).
52. Swiss Institute of Bioinformatics, "Virion", http://viralzone.expasy.org/all_by_protein/885.html
53. C. Xiao and M. G. Rossmann, "Structures of giant icosahedral eukaryotic dsDNA viruses," Current Opinion in Virology **1**, 101 (2011).
54. F. Merzel and J. C. Smith, "Is the first hydration shell of lysozyme of higher density than bulk water?" PNAS **99**, 5378 (2002).
55. T. Søndergaard and K. H. Dridi, "Energy flow in photonic crystal waveguides," Phys. Rev. B **61**, 15688 (2000).
56. M. G. Rossmann, M. C. Morais, P. G. Leiman, and W. Zhang, "Combining X-Ray Crystallography and Electron Microscopy," Structure **13**, 355.
57. M. Vulović, R. B.G. Ravelli, L. J. van Vliet, A. J. Koster, I. Lazić, U. Lücken, H. Rullgård, O. Öktem, B. Rieger, "Image formation modeling in cryo-electron microscopy," J. Struct. Biol. **183**, 19 (2013).
58. F. Fogolari, A. Brigo, and H. Molinari, "The Poisson–Boltzmann equation for biomolecular electrostatics: a tool for structural biology," J. Mol. Recognit. **15**, 377 (2002).
59. J. Gruber, A. Zawaira, R. Saunders, C. P. Barrett, and M. E. M. Noble, "Computational analyses of the surface properties of protein-protein interfaces," Acta Cryst. **D63**, 50 (2007).
60. Z. Shang and F. J. Sigworth, "Hydration-layer models for cryo-EM image simulation," J. Struct. Biol. **180**, 10 (2012).
61. H. Liu, B. K. Poon, A. J. E. M. Janssen, and P. H. Zwart, "Computation of fluctuation scattering profiles via three-dimensional Zernike polynomials," Acta Cryst. **A68**, 561 (2012).
62. V. Elser, "Strategies for processing diffraction data from randomly oriented particles," Ultramicroscopy **111**, 788 (2011).
63. N. Baddour, "Operational and convolution properties of three-dimensional Fourier transforms in spherical polar coordinates," JOSA A **27**, 2144 (2010).

# 1. Introduction

## 1.1. Context of the problem

The recent efforts seeking to extend the methodology of protein crystallography to nano-crystals and eventually single particles [1] have resulted in appropriate alternatives for the technology behind crystallography in most aspects: sample preparation, sample delivery, light source, optoelectronic detection, and statistical analysis. It seems however, that the (over-) simplified model of light propagation and its crucial implicit assumptions have been taken in many (if not most) cases for granted. This model (*Geometrical* model, elaborated in *Part 1*) is not directly applicable to arbitrary illumination schemes or arbitrary heterogeneous objects (even arbitrary crystals).

According to the Geometrical model, at large (classical) scales, the pattern of scattered light from a homogeneous object is material-independent. The pronounced signatures of morphology may confirm this prediction in many cases in a qualitative or semi-quantitative way. However, it is clear that light undergoes more distortions in one material compared to another, and hence the constraint on the validity of the model itself is material-dependent.

Furthermore, the sought 3D "density map" can have physical meaning and mathematical properties different from those of the density map in crystallography.

### 1.2. Definition of the problem

Experimental scattering patterns from symmetric nanoparticles [2] or viruses [3] have shown major signatures of morphology, consistent with the prediction of the Geometrical model. According to this model, in volumetric imaging of an object made up of a homogeneous core embedded in a homogeneous shell, the core can have a relatively weak signature (Appendix A). The key question is if the weak residual patterns have also some volumetric information; and if yes, what constraints exist for the retrieval.

### 1.3. Outline

This contribution has been structured as follows: Nontrivial aspects of single particle light scattering and hence some limits of the conventional model are exemplified in the test case of a sphere in Section 2. In Section 3, some modified schemes of the common model with more accurate description of light and the associated scattering are described. Electromagnetic boundary conditions, non-classical scattering, and different meanings of the "density map" are addressed in Sections 4, 5, and 6, respectively. Practical implications for "3D imaging" of viruses, apparent enhanced electron densities on surfaces, hybrid X-ray/Electron Microscopy reconstruction, and further complications with phase retrieval and uniqueness in single particle imaging are discussed in Section 7, and conclusions are made in Section 8. Appendices $A$, $B$, and $C$ include mathematical derivation of the materials addressed in earlier Sections.

An approximate Venn diagram of different regimes of X-ray scattering is shown in Figure 1. It is the big picture, within which the interrelation between the materials in subsequent Sections and those in *Part 1* can be seen.

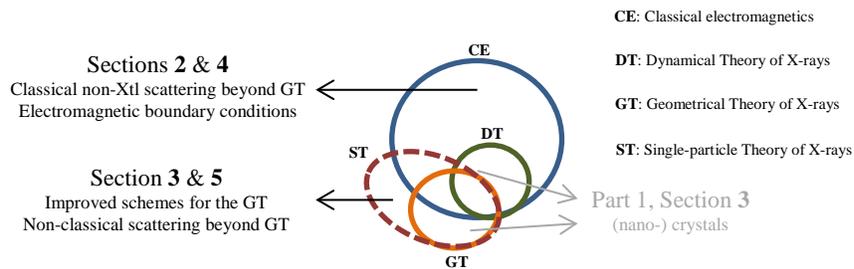

Fig 1. Venn diagram showing different regimes of X-ray scattering and the interrelation between them. It is the big picture, within which the interrelation between the materials in subsequent Sections (and also those in *Part 1* of this report, which have been grayed out here) can be seen.

## 2. Test case of a sphere

### 2.1. Approximate limits of "weak" X-ray interaction

One trivial challenge to the Geometrical model of scattering is the accumulated *extinction* (combined effect of phase shift and absorption, characterized by a complex refractive index) during the course of propagation. An intuitive condition for self-consistency is having a small difference between a plane wave having passed through an object and one having propagated in the background medium. For the case of a homogeneous sphere, this means $4\pi R|n-1| \ll \lambda$, which is known as the Rayleigh-Gans condition. Far from resonances, $|1-n|$ varies proportional to $\lambda^2$ for X-rays, and the Rayleigh-Gans condition requires $R \ll k/\lambda$, where $k$ is a constant. So, the safest regime is smallest wavelength far from resonances.

Using the correction in Sec. 3.2, the Rayleigh-Gans inequality can be converted to a real-space measure of error. If $R_c$ is the radius for which the Rayleigh-Gans formula becomes an equality $4\pi R_c|n-1| = \lambda$ (*critical radius*), then the error for considerably smaller sizes is $E = \frac{R}{R_c/50} \times 1\%$. Assuming a quadratic variation of $|1-n|$, *the error is simply proportional to $R\lambda$* for a given material and far from absorption edges.

Table 1: Refractive indices and critical radii for a homogeneous sphere made of Gold or Carbon at three typical X-ray energies. Refractive indices retrieved (with verified mass densities) from http://henke.lbl.gov/optical_constants/getdb2.html

| Element | $E_{X-ray}$ (keV) | $(1-n) \times 10^{-5}$ | $R_c$ (nm) |
|---|---|---|---|
| Carbon | 0.5 | 184+59.5i | 102 |
| Carbon | 1.8 | 14.5+0.5i | 378 |
| Carbon | 10 | 0.457+0.0005i | 2158 |
| Gold | 0.5 | 486+476i | 29 |
| Gold | 1.8 | 66.8+15i | 80 |
| Gold | 10 | 2.99+0.22i | 329 |

### 2.2. Valid (and seemingly-valid) results beyond the Rayleigh-Gans limits

In special cases, the Geometrical model may remain or look valid for big objects violating the Rayleigh-Gans inequality:

1) An empty shell affects the propagating beam only at entrance and exit surfaces, and the effective path length is much smaller than the geometrical depth. This argument is only based on accumulated extinction and discards other distortions. While the Geometrical model can be valid for volumetric reconstruction in this case, there is indeed a (filled) volume of zero, and the measurement is indeed *morphology*.

2) Extremely high extinction (say with Gold at 0.5 keV) can distort the beam considerably even with one (entrance) facet of a shell. Assuming dominant scattering from the entrance facet with a reduced effective thickness, it may be considered *1-sided morphology*.

3) If optical distortions can be considered as illuminating an *effective shell* rather than the entire volume, there are clear signatures for this difference. However, these signatures seem to be usually sacrificed in the design of scattering experiments. In the framework of the Geometrical model and within a scale factor, the scattered electric field patterns from a spherical shell and from a solid sphere look very similar: $E_{solid}(x) = [\sin(x) - x\cos(x)]/x^3$ and $E_{shell}(x) = \sin(x)/x$, where $x = 2\pi R|\mathbf{k}_{out} - \mathbf{k}_{in}|$ is the normalized size of the scattering vector. Both patterns $|E_{solid}|^2$ and $|E_{shell}|^2$ represent concentric circular rings with decaying intensity envelope and nearly-similar fringe spacing in 2D scattering patterns. The major difference between the two patterns is the absolute positions of intensity minima and maxima; especially for the very first minimum (a difference of 40%). However, many scattering experiments do not measure scattering at such small values of the scattering vector

(The intuition developed from 2D phase retrieval and assuming the Geometrical model has created confidence in successful reconstruction even with several missing fringes).

4) If a large illuminated focal volume is only filled with a few small particles (as is common with jet-based sample delivery), and the beam remains a plane wave across this volume, the Geometrical model can be applicable to the large *sparse* volume. It is noted however that the mapping from the camera coordinate to the scattering coordinate can become position-dependent.

5) In a *perfect* crystal, the effective absorption can be significantly smaller (or larger) than that of bulk; Bromann's effect [4]. The Geometrical model is accurate (at least in the sense of rocking curve for specific diffraction spots) of Si and Ge crystals for thicknesses up to 1 micron [5]. In *ideally-imperfect* protein crystals, the theory of mosaicity differentiates between the small size of a coherent subunit and the large geometrical size of the crystal (*Part 1*).

*2.3. Analytic solution of the Electromagnetic formulation*

Scattering of a plane electromagnetic wave from a sphere has a known analytic solution, credited to Lorentz-Mie. Generalization to alternative shapes such as coated- or multi-sphere system and alternative excitations, such as off-axis focused incident beam have also analytical solutions [6]. These solutions, however, are infinite series of vector spherical harmonics, the coefficients of which develop 0/0 ambiguity for large objects, or extreme values of the complex refractive index. Both these conditions apply to biomolecules and similar organic macromolecules.

Some numerically-robust schemes of implementations have been reported decades after analytical formulations, and many open-source codes come with explicit disclaimer and uncertainties about potential numerical errors. As such, they cannot be used readily for benchmarking the approximation of the Geometrical model and the signature of volumetric information in typical X-ray scattering experiments. We delay such a discussion until the completion of an ongoing work for numerically-robust versions of Mie scattering for typical X-ray scattering experiments on biomolecules. However, it is insightful to see the *trends* observed in simulations and correlations with experimental data, reported in the pioneer works of Rupp [7] and Bostedt [8] on clusters.

*2.4. Surface waves (whispering gallery modes)*

In many optical experiments with a homogeneous or a periodic object, external (and also internal) illumination is restricted to the interior of the cross section. Despite partial illumination of the *object*, the *region of interest* is fully illuminated. In general single-particle experiments, however, the region of interest is the entire object, and with ordinary reconstruction schemes should be illuminated completely. This makes surface effects inevitable. The nontrivial issue of surface effects was addressed before for the case of crystals with induced offset/asymmetry in rocking curves (*Part 1*). Here we examine surface effects for the case of a sphere.

Formulation of scattering off symmetric homogeneous objects such as spheres shows partial light propagation in the form of *resonant* surface waves. These *whispering gallery modes* can generate electric fields on the surface of the object orders of magnitude stronger than that of the beam illuminating the internal volume. In the transient picture, optical energy is partially stored in the object, and partially scattered during the initial phase of illumination. After reaching steady state, the rate of transfer of energy into and out of the object are the same, except for possible loss in bulk. Upon termination of illumination, the object may remain illuminated for quite long time, before the lossy modes decay. Considering a sphere as the extreme case of a thick lens, the dynamics of internal illumination and the long effective "lifetime" can be calculated rigorously and explained intuitively for both homogeneous [9] and coated [10] spheres.

An intuitive picture of light propagation in matter is based on analogy with (the simplified zero-dimensional) electric circuits and the delivery of electromagnetic energy to a resistor. With surface modes, this static picture is changed to a dynamic one, and the effective *impedance* of the medium develops a reactive (imaginary) component. This dynamic property of the medium can introduce significant changes, even in the steady-state behavior with continuous-wave (CW) excitations. With pulsed laser excitations, the transient behavior can also come into the picture and introduce more nontrivial behaviors (compared to a medium with instantaneous optical response).

Coupling of external illumination to such surface modes can be enhanced by orders of magnitude, with focused off-axis illumination; i.e., optimizing the overlap between the incident and resonant electric field profiles. The highest intensity of scattered light from a sphere can correspond to such an off-axis illumination scenario with more ehnaced signatures of morphology rather than illuminating the volume. In other words, misleading feedback for intuitive optical alignment by increasing "the signal".

In the geometrical limit (size >> wavelength), the extinction cross section of a sphere is twice its expected value from Geometrcial model. This "paradox" originates from the exssistence of surface waves; a wave phenomenon at arbitrary scales [11]. The enhancement, the localization, and the delay of such surface waves can have important implications. Pumping and depleting a sphere with optical energy is a dynamic process. In the presence of a dynamic damage mechanism, a measured scattering pattern corresponds to the spatio-temporal evolution of two coupled mechanisms (classical electromagnetics and damage). It is especially important with common non-gated and long-exposure-time detectors.

## 3. Improved schemes for the Geometrical model

The level of improvement provided by the following schemes is beyond the scope of this contribution. They simply make a formulation based on the common Geometrical model more self-consistent and/or robust.

### 3.1. Full 3D non-singular formulation of focused beams

Most X-ray scattering experiments use focused beams to localize and to enhance the coupling of light to a sample. Focusing X-rays with wavelengths smaller than ~2nm and focal spots greater than ~100nm is not tight-focusing, per se. However, the ultimate positioning control (actuator step size, hysteresis, pointing stability of light source, local curvature of phase front, finite width of the sample-containing liquid jet …) puts some limits on the accuracy of plane wave or even the more accurate spherical wave [12-14] approximations.

Paraxial propagation of light in free-space is characterized with a series of orthogonal electromagnetic modes, with the fundamental being a *Gaussian beam*. A light source with structured beam profile can be modeled with a few orders of orthogonal paraxial modes. However, the emission of a high-quality coherent light source (laser) is dominantly a Gaussian beam. The $M^2$ figure of merit in laser specification sheets quantifies the accuracy of this assumption and the quality of the spatial profile, based on the far-field divergence angle of the laser beam. Also, an important property of a lens is converting a Gaussian beam to another Gaussian beam.

Contrary to the spherical wave model, the Gaussian beam features a smooth diffraction-limited waist and not a singularity. Also, in addition to the finite-curvature asymptotically-quadratic transverse phase, it features an accumulated longitudinal phase shift of $\pi$ required to satisfy Maxwel's equations (Gouy phase shift). By defining the auxiliary variables $w(z) = w_0\sqrt{1 + (z/z_R)^2}$, $z_R = \pi w_0^2/\lambda$, $R(z) = z[1 + (z_R/z)^2]$, and $\zeta(z) = \tan^{-1}(z/z_R)$, the electric field of the Gaussian beam in free-space (excluding the trivial propagation phase shift, which is accounted for by the Fourier kernel in Geometrical model), is written as

$$E(r,z) = E_0 \frac{w_0}{w(z)} \exp\left[-\frac{r^2}{w^2(z)} - \frac{i2\pi r^2}{2\lambda R(z)} + i\zeta(z)\right]$$

Figure 2 contrasts the effects of the phase profiles of the Gaussian beam and the spherical beam models of focused illumination simulated with the Geometrical model of scattering for a sphere with a diameter of $300nm$ illuminated with a beam with intensity full-width-half-maximum width of $100nm$ ($w_0 = 72nm$), energy of $512eV$ ($\lambda = 2.4nm$) and with longitudinal and transverse offsets of $z = 140nm$ and $r = 20nm$, respectively.

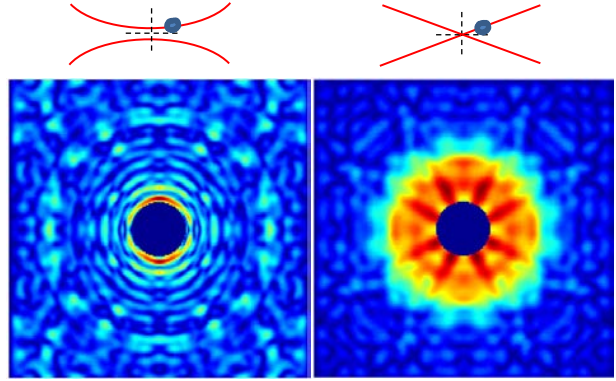

Fig. 2. Simulated scattering (Geometrical model) of a sphere excited with a focused beam with an intensity full-width-half-maximum width of 100nm and photon energy of 512eV with an offset of 20nm/140nm along the transverse/longitudinal directions. The left (right) panel corresponds to the 3D Gaussian (spherical wave) model.

*3.2. Local Field estimation (extension of Beer-Lambert law to a heterogeneous medium)*

The Geometrical model can be made more self-consistent by explicit consideration of the accumulated extinction. This *scalar* model still discards reflections and distortions to the phase front. The extinction is assumed to be locally *smooth* and globally not very significant. With weak interactions across infinitesimally-thin longitudinal slices and using what may be considered a generalized form of Beer-Lambert's law, the following expression for the projected density on the camera can be derived, as detailed in Appendix B:

$$\boldsymbol{E(x,y) = \left(1 - \frac{i\pi}{\lambda_0}E_{ideal}\right)E_{ideal}}$$

As the equation implies, the density itself has no physical dimension, but its projection gains the dimension of length and unit of $m$ in the SI system. A measured snapshot is proportional to the magnitude squared of the 2D Fourier transform of $E(x,y)$. The error of a thick or dense object appears as a self-modulation term. The pixels corresponding to the periphery of an object (with strong signatures in Fourier transform) have negligible distortion. On the other hand, innermost voxels with large values of projections undergo large distortions, too. This can make 3D scattering even more representative of morphology (rather than volumetry) in a spurious way. By combining the above self-modulation error term $i\pi k_0 E_{ideal}$ and the critical Rayleigh-Gans radius for a sphere $R_c = \frac{\lambda}{4\pi|n-1|}$, the *maximum self-modulation error* for a sphere can be written as

$$Error = (\pi/\lambda_0)(2R)\,|\delta n| = \frac{R}{R_c/50} \times 1\%$$

This correction scheme not only helps to translate the Rayleigh-Gans inequality into more tangible and quantitative terms, but also provides correction schemes for the cases where it is (slightly) violated, and also generalizes it to more general case of heterogeneous objects.

Such a real-space measure of error seems to have a better affinity with the sought real-space density profile, compared to a scattering-space measure of error. A small inclusion in an otherwise-homogeneous medium can result in only slight decrease of fringe visibility in scattering patterns. In the case of spherically-averaged scattering (SAXS), such minor differences in the scattering space may even correspond to significant real-space morphological differences between two solid objects with the same volume [15].

The Geometrical model (with many of its key assumptions) is still at the core of this hybrid formulation. Sustainability of the phase front (or lack thereof) can be viewed from the perspective of paraxial propagation. The $\vec{k}$ vector of propagation is normal to phase front, and (according to generalized Snell's law for a bent ray,) its slight variation is normal to the gradient of refractive index density $\delta\vec{k}.\nabla n = 0$. Phase front is more likely to be sustained if variations of refractive index happen transversally and not longitudinally. The extreme case of purely-transversal $\nabla n$ (longitudinally-uniform density) can be a waveguide and support guided electromagnetic modes [16]. A guided mode has a finite width (contrary to an ideal plane wave), yet undergoes no diffraction (contrary to a Gaussian beam).

*3.3. Disentangling the contributions of core and shell*

Many biological samples used in single-particle experiments [3], such as double-stranded DNA viruses [17] have stable solid-phase morphology, yet a viscous inclusion inside which may change its shape (or even topology) from one snapshot to another. Given this important challenge and the aforementioned pronounced signature of morphology in scattering patterns, there is a need for disentangling the signatures of morphology and volumetry in measured scattering patterns.

For simplicity, we consider the case of spherical geometry, yet arbitrary densities for core-shell. The geometry and the density of a given region (the core or the shell) are qualitative notions and are not quantitatively unique or independent. An icosahedral shell with uniform density, for instance, may be considered as a spherical shell with non-uniform density. So, the assumption of spherical boundaries for core and shell is not that restrictive.

The key ideas in our formulation are
- 3D formulation of real-space density with linear 1D parameterization: $\{P\}/\{Q\}$ coefficients for shell/core
- Analytical calculation of 3D Fourier transform as a linear operator on $\{P\},\{Q\}$
- Projecting *phased* measured snapshots on two (mutually) orthogonal bases (say sine and cosine): $\{C\},\{D\}$
- Deriving closed-form linear equation for calculating $\{P\}/\{Q\}$ in terms of $\{C\},\{D\}$

The final result of the derivation, detailed in Appendix C, is the following simple linear matrix equation:

$$\begin{bmatrix} \alpha & \beta \\ \delta & \omega \end{bmatrix} \begin{bmatrix} P \\ Q \end{bmatrix} = \begin{bmatrix} C \\ D \end{bmatrix}$$

This is a linear equation to solve $\begin{bmatrix} P \\ Q \end{bmatrix}$ (real-space radial weights of the reconstructed density) from $\begin{bmatrix} C \\ D \end{bmatrix}$ (reciprocal-space radial weights of the *phased* measured snapshot). The elements of the matrix $\begin{bmatrix} \alpha & \beta \\ \delta & \omega \end{bmatrix}$ are simply the *constant* projections of *known* basis functions on each other. They depend on the reconstruction parameters $R_1, R_2, q_{max}$, but are independent from measured data.

An important simplifying assumption here is taking phase retrieval for granted. This assumption is questionable, especially in the case of truncated scattering patterns (Sec. 7.5).

Nevertheless, the formulation here serves as a starting point for more elaborate ones that take the issue of the loss of phase and (the uniqueness of) its retrieval into consideration.

*3.4. Fundamental shortcomings of the Geometrical model*

In brief, the followings are some important issues associated with the Geometrical Model:

- Attributing bright-dark fringes of a homogeneous object to mere geometrical parameters
- Missing the existence and variations of excess cross section
- Missing long-lived circulating resonances with pulse excitation
- Missing the significant impact of boundary conditions

## 4. Electromagnetic boundary conditions

The impact of an obstacle such as a stone on the normal flow of water in a river is familiar patterns and vortices that conserve the continuity of liquid flow. Similarly, the flow of electromagnetic energy maintains the continuity of (tangential) electric and magnetic fields across a boundary. In the case of a plane wave crossing the plane boundary between two bulk media, these boundary conditions give rise to the conservation of the tangential component of $\vec{k}$ vector across the interface; i.e., Snell's law of refraction. Nontrivial propagation of X-rays in perfect crystals and unorthodox rocking curves, or nontrivial damage of a homogeneous object starting at the back side addressed in *Part 1*, originate from the compliance of light to these boundary conditions.

*4.1. Diffraction tomography*

To best of our knowledge, the only single-particle reconstruction beyond the Geometrical model is based on diffraction tomography [18]. It includes higher-order perturbative solutions of scalar Helmholtz equation (higher-order Born's approximation). While *a meaningful and major step forward*, it is still based on the assumption of weak light-matter interaction and misses the reflections at grazing incidence angles, for instance.

The key point is an initial approximation of a homogeneous medium and using its characteristic response (Green's function) for the entire *inhomogeneous* medium with unknown boundaries [19]. The key idea behind Diffraction Tomography can be used in a more rigorous way by explicit consideration of electromagnetic boundary conditions in a heterogeneous object composed of multiple homogeneous units with arbitrary shapes and indices [20-22].

Electromagnetic boundary conditions are considered by the Electromagnetic and the *perturbative* Hybrid models, but not by the Geometrical model (w/wo corrections).

*4.2. Geometrical model implies a "hard" boundary*

Illuminations confined to the interior (and avoiding the periphery) of a homogeneous or a periodic object can also generate nearly zero intensity (internal illumination) on the edges of the transverse boundary. With zero intensity, one can model light propagation by imposing almost any boundary condition (periodic, perfect electric conductor …) on this transverse boundary. However, the actual physical boundary condition, applicable to arbitrary illuminations (relevant to nanocrystals and single particles), depends on material properties among other things.

A crystal with a transverse cross section larger than the beam diameter supports electromagnetic *modes* [23,4] that under appropriate conditions reduce to a single plane wave. This plane wave (and hence the validity of the Geometrical model) originates from not only weak light-matter interaction; but also from the *existence and the special form of electromagnetic mode* in a crystal and *exciting it with an appropriately focused/collimated*

*beam.* In principle, this luxury of sustained modes propagating in volume does not exist in the case of *volumetric* single-particle imaging, even with weak light-matter interaction (small object or hard X-rays).

With Geometrical model, one ends up with a shape transform (*Part 1*; Sec. 3) dependent merely on geometry and not material properties. This *signature* is similar to the signatures of hard boundary conditions (vanishing electric or magnetic fields on interfaces) by metallic ($n = i\infty$) or strong magnetic ($\mu_r = \infty$) shells, and not in (possibly lossy) dielectric bulk [16]. The distinction between these boundary conditions becomes clearer by contrasting the dependence of light propagation constant (effective 1D k-vector) in these two cases (Figure 3).

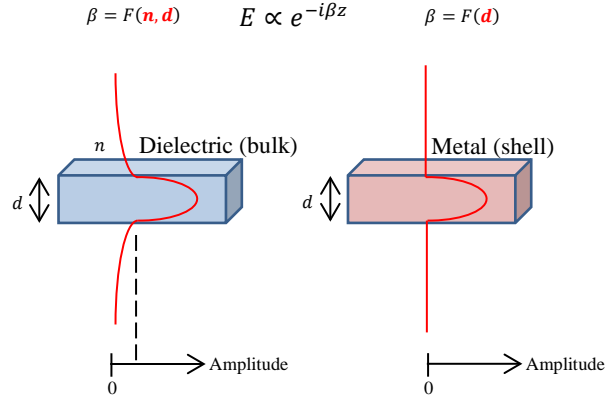

Fig. 3. Electric field patterns of the fundamental mode of a dielectric (core) and a metallic (shell) waveguide. The propagation constant of the mode in a hollow metallic waveguide (at a given frequency) depends only on the geometry, whereas that of a dielectric waveguide depends on both the geometry and material properties. The "hard" boundary condition in the case of the metallic shell is similar to that of the Geometrical model.

In the case of an incident beam hitting the edges of an object, the Geometrical model seems to introduce a spurious boundary condition on an object, even a crystal [5], immersed in the incident light. This effect may be negligible for an object with relatively large homogeneous zones (such as binary core-shell with smooth surfaces), but becomes more important for an object with a high level of heterogeneity and/or surface roughness.

*4.3. Polarization*

To a laser operator, polarization is a property of *light source*; $s/p$ polarizations corresponding to the electric field perpendicular to / parallel with the floor (for a well-aligned laser beam). However, polarization, as relevant to scattering, is defined with respect to the *entrance facet of an object* (over which boundary conditions are satisfied by including the incident beam). For crystals under "usual" illumination schemes, special symmetric objects, and also for 1D or 2D problems, there exist two decoupled polarizations. It means that finding the 6 unknown components of the electric and magnetic fields is split into two independent problems (with three field components being zero in each case). Under such conditions and with good alignment, the two polarizations relevant to scattering are related to the laser operator's $s/p$ polarizations in a trivial way.

In the general 3D case, the notion of decoupled polarizations does not exist in the first place. Excitations with $s/p$ polarizations provide complementary, but not decoupled information (of diagonal and off-diagonal elements of the scattering matrix). Weak index contrast of X-rays and working at small scattering angles can introduce considerable simplifications. Polarization factors developed in the context of crystallography may still be good approximations for single particles. We are not aware of the scope of validity of such an approximation, especially for large scattering angles (where it matters).

In the "Dynamical Theory" of X-rays, polarization is defined with respect to the displacement vector $D$ and not the electric field vector $E$ [4].

## 5. Non-classical scattering

### 5.1. Different regimes of light-matter interaction

Classical scattering in an object made up of (linear non-magnetic) *bulk* subsets is governed by spatial distribution of the refractive index. A plane wave is scattered if only it reaches the interface between two bulk media. The propagation constant of a plane wave (even if complex) is preserved, and no new k-vectors are generated. This notion is independent of the size/wavelength ratio. This classical picture can fade out at wavelengths smaller than ~40nm or scales smaller than ~10nm, as shown in Figure 4. The former is observed in non-classical (UV or) X-ray scattering from bulk at resolutions better than $\sin(\theta)/\lambda \sim 1/(40nm)$. The latter originates from quantum confinement and modifies the shape and the spread of electron density and hence light-matter interaction in the $< \sim 10nm$ scale.

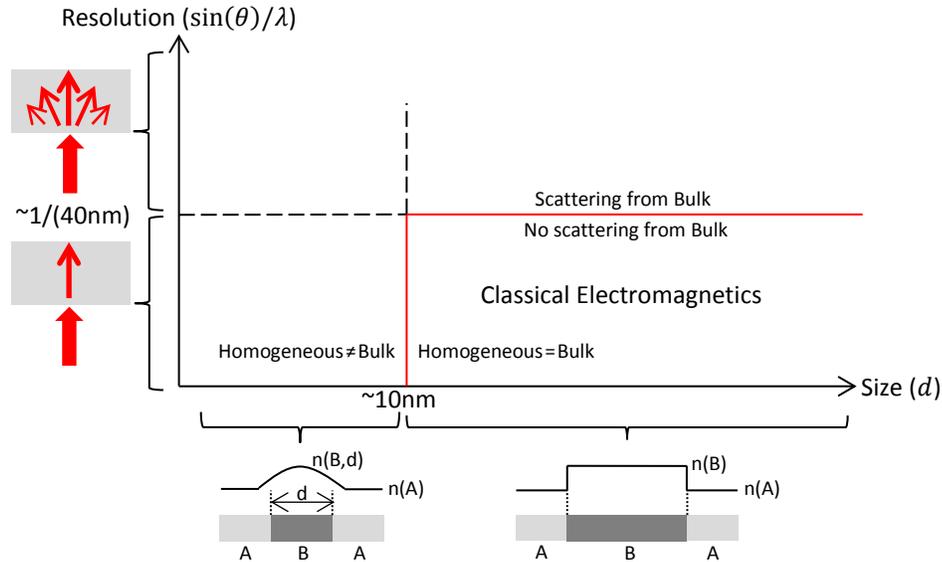

Fig. 4. The boundaries of the notion of *bulk* and *direct* applications of classical electromagnetics: The cartoons underneath the horizontal axis show that at small scales, a chemically-homogeneous material does not necessarily have a uniform or even confined electron cloud. The cartoons along the vertical axis show that at high resolutions, a plane wave is not a sustained electromagnetic mode of a homogeneous medium and causes scattering.

The majority of X-ray scattering models have been applied and validated in the contexts of 1) large optical elements, in which case the notion of bulk is *revived*, or 2) ideally-imperfect (protein) or perfect (semiconductor) crystals with possible perturbation, in which case the notion of bulk is not required. Applying classical electromagnetics to other unexplored regimes, such as $< 10nm$ ordered (homogeneous, periodic, cluster) media, should be done with care. An insightful test case for 3D single-particle experiments can be 1D metrology of few-nanometer films and the comparative performance of X-ray measurements [24].

*5.2. Semi-classical models at small scales (d < ~10nm)*

The correspondence between the density profile and a constituent homogeneous unit can be lost because of quantum confinement at $< \sim 10nm$ scale. Furthermore, the dynamics and the propagation of light (especially with intense laser pulses) can be coupled with those of the quantum energy levels and can create nontrivial phenomena [25].

Potential signatures of this regime of light-matter interaction are briefly reviewed in the optical (VIS-NIR) and less-explored X-ray regimes, and two models of effective index and effective polarization are mentioned. Finally, the regimes of validities of different models are *roughly* visualized.

5.2.1. Effective index model at optical (VIS-NIR) wavelengths

As a first approximation, light propagation in an ordered region with a size smaller than $\sim 10nm$ may be approximated as having a modified refractive index (different from that of bulk). In microelectronics fabrication technology, this assumption has been used for multimodal metrology of ultra-thin films. At larger scales, an optical technique (ellipsometry) can use the refractive index of bulk to measure the film thickness. At small scales, however, the thickness itself can be first determined using the electrical tunneling effect. Ellipsometry can then be used in a mode opposite to the conventional way to extract the effective refractive index from the known thickness [26].

For a thin $SiO_2$ film grown on a Si substrate and using the data from 4-6nm films, the refractive index (at 633nm) has been modeled phenomenologically as a Gaussian function of thickness [26]. Full complex dielectric constant has also been quantified and compared with that of bulk in the case of silicon nanocrystals embedded in $SiO_2$ [27]. A more rigorous approach to size-dependent dielectric function has been formulated in the context of scattering from spherical particles to overcome the deficiency of *classical* Mie-scattering [28].

5.2.2. Impact on X-ray signatures

Soft X-ray absorption edges and shapes change considerably in going from bulk silicon to *nanocrystal* silicon [29-31]. Absorption itself is a contributor to the optical density and scattering. Furthermore, it is correlated with the refractive index via the Kramers-Kronig relation. A change in the absorption spectrum also implies a change in the dispersion (the spectrum of the refractive index).

In the classical picture, the total extinction (absorption and scattering) of a homogeneous sphere depends critically on the size of the object for sizes on the order of the wavelength [11]. This effect should be disentangled from potential quantum confinements in objects with $< \sim 10nm$ size [28].

Such potential quantum confinement signatures have also been studied in the context of clusters [32], and differences between experimental and theoretical results have been reported. The theoretical results seem to have been purely quantum mechanical calculations of energy levels with no consideration of (nontrivial) light propagation.

In the case of 1D ultra-thin film measurements, reflectometry with grazing-incidence X-rays and neutrons had been anticipated to give the highest accuracy. Despite acceptable results, the high hopes for accuracy have not come true. The exact reason is not clear; yet, it is thought to have contributions from the refractive index [24].

5.2.3. Effective polarization model: coupled evolution of optical and electronic waves

Aside from quantitative errors, the effective index approach may even fall short of a qualitative story by missing important non-classical signatures of light-matter interaction (especially in excitation with a laser pulse, such as self-induced transparency). A better approximation in the case of a quantum-mechanical N-level system is to include a semi-classical induced polarization dependent on both the classical electric field and the quantum mechanical densities of energy levels. The *semi-classical Maxwell-Bloch model* is commonly used for modeling the coupled generation and propagation of light either at optical [33,34] or X-ray wavelengths [35-37]. This model seems to be the best candidate for capturing the *coupling of optical propagation and quantum confinement* in ordered low-dimensional objects, especially with intense pulse excitations. In terms of numerical considerations for suppressing the error of finite-difference approximation, special schemes for these nonlinear systems [25] beyond the conventional ones in the classical linear framework [38] are necessary.

5.2.4. Rough sketch of regimes of validity of different models

The materials presented so far in terms of the limits of the Geometrical model and more accurate alternatives can be represented in a multi-dimensional space with rough boundaries. To facilitate the visualization, we restrict the dimensions to two, with each axis corresponding to a "regime" and representing more than one parameter. The result is illustrated in Figure 5.

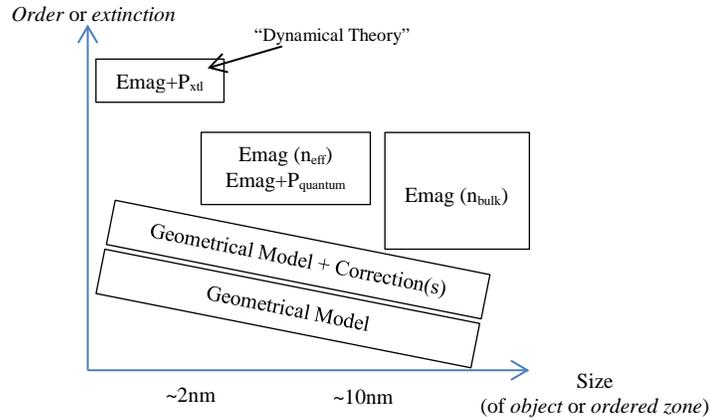

Fig. 5. Representing different *(semi-) classical* models of X-ray scattering with rough sketch of their regimes of validity

5.3. Semi-classical models at high resolutions $(sin(\theta)/\lambda > \sim 1/(40nm))$

In the Electromagnetic picture, the polarization $\vec{P}$ induced in homogeneous isotropic bulk in response to the electric field $\vec{E}$ of incident light is proportional to and in the direction of the electric field of light $\vec{P} = \epsilon_0(1 + \chi)\vec{E}$. As the wavelength decreases (below $\sim 40nm$), the (UV or) X-ray photons cause a classical acceleration of the local electron density, which in turn radiates according to dipole-like radiation (similar to Thomson scattering). The continuum of scattered k-vectors in response to the incident k vector implies the failure of the notion of *bulk*, and a plane wave being its *mode* (eigenfunction). We consider three different approaches in handling this effect:

    1. With Geometrical model (w/wo corrections), as in conventional protein crystallography, one does not need a classical assumption (This tangible convenience, however, comes at the cost of making classical arguments in *Part 1* less rigorous). Whether the scattered light (up to very large angles and resolutions) originates classically or non-classically is only reflected in the value of the form factor [39], without changing formulations.

    2. With classical electromagnetic formulations based on the notion of bulk (such as Mie scattering), one should limit the generated scattering to small enough resolutions: $\sin(\theta_{Max})/\lambda < \sim 1/(40nm)$. The limit $40nm$ comes from the smallest photon energy, for which non-

classical scattering has been tabulated. Alternatively, one may use a threshold value of $\sin(\theta)/\lambda$, for which the form factor of Carbon [39] drops by less than 2%, for instance. This accommodated non-classical error can be used to define a measure of self-consistency of a classical formulation.

3. The most rigorous way of treating non-classical scattering is introducing an effective induced polarization. This has already been "done" (or in a sense bypassed) in the very special case of a perfect crystal [40]. This formulation has two aspects: A) The assumption of a periodic polarizablity (or refractive index), and B) A classical treatment of light propagation in such a medium. *Irrespective of how light would propagate in a homogeneous medium, it is reasonable to assume that it develops a periodic envelope (Bloch wave) in a crystal*. This behavior can be modeled (semi-) classically by an effective periodic refractive index (which is not limited to macroscopic scales anymore). However, the unanswered question of propagation in a homogeneous medium is exactly what matters to single particle scattering; differentiates it from perfect crystal scattering; and requires a non-classical formulation of the induced polarization.

The induced polarization has been used to quantify the coupled transmission, reflection, and scattering in the cases of anisotropic crystals and static nonlinearities [41], dynamic nonlinearities [42], and N-level quantum systems [25]. In all these cases, however, the induced polarization at a point is defined with the total electric field at that point, only. In the case of X-ray scattering, Thomson-like scattering couples the density and illumination gradients in a nontrivial way.

One way to get around this difficulty is to skip the notion of *bulk*, and if possible to include the notion of form factor. An important question in electromagnetics is if and under what conditions, the notion of bulk and the notion of oriented dipoles are equivalent. The consistency of these two models requires working at "large" wavelengths (compared to the size of dipoles) and a correction to the electrostatic (Clausius–Mossotti) term. This notion has been used in a model of classical far-field scattering, known as the discrete dipole approximation or DDA [43]. It considers the total local electric field to be the sum of the incident beam and scatterings from all other dipoles. It uses the accurate dipole scattering, which only in the far-field approximation is reduced to Thomson scattering.

We anticipate that *the DDA method with small enough dipoles, yet with no need to complying with the notion of bulk*, can be used to model the *coupled illumination and scattering*. This issue is currently under further investigation. From a computational point of view [44], the DDA method reportedly outperforms the more common T-matrix method (calculating the scattering of a system from the scatterings of individual components) [45]. It also enjoys FFT-based accelerated computation [43].

## 6. Meanings and properties of "Density map"

### 6.1. Optical vs. electronic vs. chemical images

The corrections for the Geometrical model elaborated in this contribution are mostly aimed at disentangling the signatures of the *optical* density (scattering potential) from the self-induced pattern that illuminates it. It is common in the X-ray community, however, to correlate the scattering patterns with *electron* density. This is a reasonable notion in protein crystallography because of 1) heterogeneity of biomolecules and 2) atomic-level resolution, in which case both *electron* density and *optical* density are (within the achievable and required resolution) practically the same clouds localized around atoms. A library of element-specific scattering patterns is used for Model Building by locally fitting the measured *optical* density [46]. The optimal fit; i.e., types and positions of individual atoms forms the *chemical* image, usually shown superimposed on the electron density image.

An important convenience in protein crystallography is the simple correspondence between (overlap of) the three optical, electronic, and chemical images. In single particle imaging, this luxury is questionable.

### 6.2. Long-range synergy of electron cloud in protein crystallography

Electron density $ED$ is related to the elctronic wavefunction $\psi$ as $ED \propto |\psi|^2$. The information content of the wavefunction for two or more bonded atoms concerns 1) their types and equilibirium positions, and 2) modulations of individual spherically-symmetric electron clouds of hypothetical isolated atoms, after forming the bond. This latter piece of information is discarded in protein crystallography

Long-range synergy of electron cloud (coherent *electronic* interactions) in biomolecules can be present even at distances up to ~10nm, as evidenced by FRET-like spectroscopic processes. Such interactions are obviously stronger at ~ 2 orders of magnitude smaller distances; i.e., the length of covalent bonds. However, fitting the measured optical density map of a *hetrogeneous* biomolecule to a chemical image can be done efficiently by discrading variations as between a $p$ orbital or two interpenetrated $s$ orbitals. The size of the *primary units* of a *hetrogeneous* biomolecule (individual atoms) is on the order of or even smaller than the achievable and required resolution distance.

*Partial coherence* in protein crystallography concerns not only *optical* waves (intensity-wise addition of scattering patterns from different mosaic blocks $I = \sum|E_{mosaic\ block}|^2$), but also *electronic* wavefunctions (superposition of electron densities $ED = \sum|\psi_{isolated\ atom}|^2$).

### 6.3. Electron cloud distortion at intermediate scales: Imaging meets spectroscopy

In an object with one few-nanometer subunit, the chemical image (*ionic body*) can have atomically sharp boundaries, and yet the electron cloud can have smooth extensions beyond the boundaries (*spill-out* effect) and (*Friedel*) oscillations within the homogeneous zone. This *soft boundary* is more pronounced in the case of metallic or semiconductor nanoparticles [28].

Such quantum confinement issues are irrelevant at low (classical) resolutions. They also seem to be irrelevant at very high (atomic) resolution imaging of biomolecules, because of hetrogeneity and very small primary units. However, in few-nanometer nanoparticles or core-shell objects, an *ordered* (homogeneous, periodic, cluster) primary unit has a size well above the resolution distance. The strength and the relevance of quantum confinement vs. size have been shown with the rough semi-quantitative trends in Figure 6.

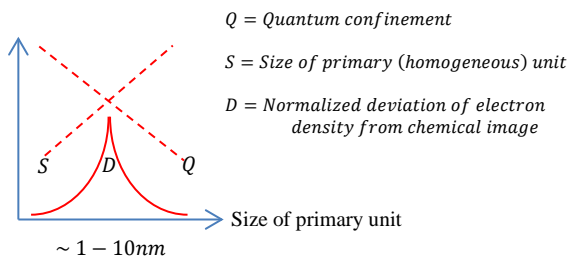

Fig. 6. Rough semi-quantitative sketch of the strength and relevance of quantum confinement and long-range synergies (coherent superposition) of electronic wavefunction.

### 6.4. Are high-frequency ripples in electron cloud spurious?

The aforementioned internal oscillations and also the spill-out of the electron cloud within and around a homogeneous medium cause authentic features in the *eletcron* and most likely in the *optical* density, but not in the *chemical* image. The existence or level of such authentic, yet visually-not-so-appealing oscillations can also be affected with reconstruction parameters affecting resolution and numerical errors of Fouier analysis on a discrete grid (similar to "B-factor" in crystallography).

In the case of spill-out effect of the electron cloud in a thin nanostructure, even a correct chemical image (equivalent of a PDB file in protein crystallography) cannot generate the electron density image accurately.

*6.5. Electron density image vs. chemical image: which one generates the other?*

"In a high-resolution electron density map, corresponding chemical residues *show up*". Is this true notion for the end-users of protein crystallography software also true for software developers of single particle imaging?

Electron density image is obtained by fitting a measured optical density image to an assumed chemical image, as shown in Figure 7. The (semi-) classical expression correlating optical density to electron densitie*s* is based on form factors and dispersion-correction terms of *specific known elements* [47,46] in primary units. Talking about "density" gradients only makes sense in the classical regime. In the non-classical regime, explicit knowledge of (or assumption about) the elements associated with a sought density are required. Different elements introduce different levels of non-classical scattering, and the combined effect of {density, form factor} determines the scattering.

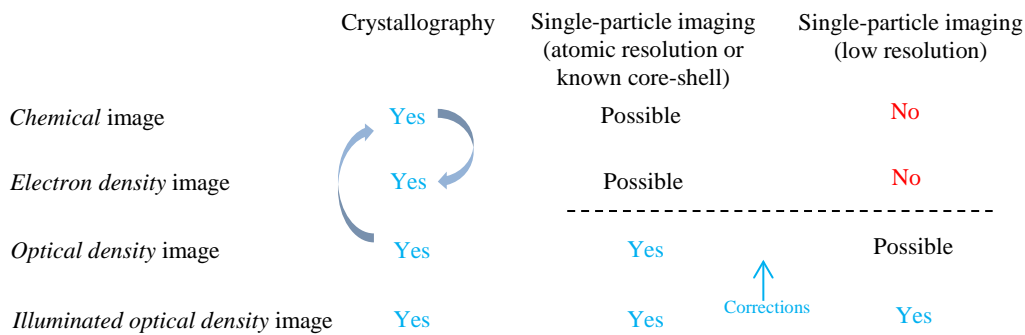

Fig 7. Meanings of the "density map" in different X-ray scattering experiments. The correction schemes reported in this contribution merely aim at getting an undistorted *optical density* image. The trivial equivalence of all four images (and hence the blurred boundary between them) in conventional protein crystallography originates from *atomic-resolution fitting* of data to known scattering signatures and heterogeneity of biomolecules. One fits the optical image to a *chemical image, from which an electron-density image is inferred* (and not the other way around).

*6.6. Optical density: real or complex?*

As seen in Table 1, the refractive index can be well approximated by a real number, as photon energy approaches higher values (and far from resonances). The same trend holds for the theoretical optical density (without corrections), which is related to electron density with a *real* proportionality constant. This is why "the density" is intuitively a real quantity in protein crystallography.

At soft X-ray energies, as used with many single particle experiments, the real-space optical density is a complex quantity:
- For a filled (topologically *simply-connected*) homogeneous object, or a homogeneous object with voids *in vacuum*, the complex refractive index (or optical density amplitude) can be factored out. Within this complex constant coefficient, the 3D map can be modeled as real.
- For a heterogeneous object or a homogeneous one with voids (topologically *multiply-connected* or *disconnected*) embedded in a background medium, the real-space density map is a complex function. It cannot be converted to a real one by simply factoring out a constant complex factor. Phase retrieval algorithms or reconstructions based on fit to a real-space profile [48] should consider the possibility of a complex density map.

At classical resolution (no contribution from form factor) and in the case of a homogeneous object without voids, one can simply consider the absolute value of the retrieved *optical* density map to also represent *electron* density map within a scale factor [49]. In other cases, the optical and electronic density maps are not necessarily the same. Estimation of electron density requires an assumption about the chemical image, and fitting the optical density map to such a model of chemical image.

## 7. Discussions and open questions

### 7.1. Practical implications for "3D imaging" of viruses

Given the strong scattering with big objects and at small wavelengths, many data sets have already been collected with big viruses and/or at X-ray photon energies close to 500eV. The gain in signal to noise ratio in such data sets comes at the cost of the compromised accuracy of the conventional Geometrical model (Table 1). But even in such cases, the clear signature of morphology [3] is a promising factor. The high extinction at 500eV may also come as a bonus *concealing rather than distorting volumetric information*. Schemes for disentangling the signatures of morphology and volumetry are expected to bring such datasets to more conclusive reconstructions of morphology. The scheme presented in Sec. 5.3 is just a starting point for developing more elaborate formulations to retrieve the capsid morphology. If volumetric distortions are not dramatic, a correction scheme, as in Sec. 3.2, may be helpful in even retrieving the signature of the genome.

A morphology-oriented (rather than volumetry-oriented) approach can be compared with a similar one in the parallel field of electron microscopy, where some recent reconstructions of big viruses have been focused essentially on the capsid (morphology). It is not clear to us whether this limitation has been caused merely by icosahedral [50] or 5-fold [51] rotational averaging of the density map (reasonable for capsid, but not for the genome inside), or possibly other factors. Considering the results of Sec. 5.2., it is expected that irrespective of the illumination source (photon or electron), the edges of the projected density experience the least self-modulation error.

Aside from corrections in modeling and data analysis, more recent experiments (for instance, LCLS proposal LB33) have also successfully measured scattering patterns of smaller viruses (PBCV-1 with a 3D capsid diameter of 190nm and a packed genome, as opposed to Mimi virus with a diameter of 750nm and loosely-packed genome) at the high soft-Xray photon energy of 1.8keV. These data sets are currently being analyzed.

Contrary to protein crystals with partial coherence and >10nm unitcells, viral capsids are decorated with seemingly more-ordered *2D* arrays of hexamers [52] or trimers [53] of smaller few-nanometer capsid proteins. Quantum confinement (Sections 6.1 and 8) in these cases, while minor, can be relatively stronger compared to that in the case of protein crystals.

On a final note, showing 2D projections without explicit specification can be misleading. A 2D projection with a structured pattern inside may correspond to the projection of an empty capsid with modulated shell or a capsid with the genome inside. This crucial single bit of information (emptiness of the capsid) can get a wrong answer with insufficiently explained 2D projections.

### 7.2. Geometrical model is (more general and) not equivalent to Born's approximation

The Rayleigh-Gans and the first-order Born's approximations *end up with a similar formulation* as the Geometrical model; i.e., the 3D Fourier transform of the density function. However, they are based on the extra assumption of classical scattering from (possibly heterogeneous, yet) *locally-isotropic* bulk. The Geometrical model, as used in protein crystallography with ultimate non-classical resolution, does not require classical concepts.

*7.3. On "enhanced electron densities on surfaces"*

Surface modes correspond to the poles of the dispersion equation, and depend on the size and the refractive index. They do exist even with a purely real refractive index (at a frequency far from resonances). Approaching the resonance frequency of bulk matter further enhances such effects or the specific near-field pattern in the vicinity of the object. But even in this case, with enhanced classical *oscillations* of the electron cloud, the *average position* of the electron cloud and the value of the refractive index remain the same (no change of bulk). Classical Mie scattering analyses of such effects cannot be interpreted non-classically with the Geometrical model (correlating illumination gradients to displaced charges). Even close to resonance frequencies, *surface waves do not imply (steady state) charge displacement.*

Quantum mechanically, one may expect carrier confinement in ultra-thin shells or similar geometries. Even if relevant to a structure, this property cannot be deduced from classical Mie scattering analysis.

Also from this perspective, we suggest further inquiries into the observed "enhanced electron density of the hydration shell" surrounding biomolecules [54]. Enhanced illumination and enhanced electron density may both have potential partial contributions to this phenomenon. Global surface waves are avoided in protein crystallography by limiting the illumination to the middle of the object. Individual mosaic blocks or unit cells, however, are likely to experience such effects locally. For guided waves in 2D crystals, a uni-directional flow of the total power across unit cells with a single optical mode (despite counter-propagating electric field components) is achieved by forming power vortices [55]. Such flux vortices, if any, appear as spurious enhanced electron density with the Geometrical model.

*7.4. Feedback from density map to solvent properties*

Making a biological sample and especially a crystal in a stable and biologically-"happy" form includes a strong constraint (optimizing parameters such as temperature and PH level of the solvent). Imaging the sample is also about a strong constraint to narrow down the large space of candidate solutions to a small one and eventually to the right one. What connects these two optimizations is the electrostatic potential on the surface of the sample:

$$\rho_{protein} \rightarrow V_{surf} \leftarrow PH_{solvent} \ \& \ T_{solvent}$$

In a hybrid electron microscopy (EM) and X-ray reconstruction [56], this dependence can be used when the resolutions of the two data sets become comparable. Note that the *electrostatic potential* (and not electron density) is the physical quantity or "density map" measured by electron microscopy [57].

Volumetric electrostatic potential map associated with an electron density is formulated quantum mechanically, and is eventually reduced to a simple projection in pure EM studies [57]. Morphological electrostatic potential on the surface, however, can be solved classically, and derived from X-ray crystallographic data. The physical constraint correlating EM and X-ray density maps on the surface is Poisson-Boltzmann equation [58]; an extention of Poisson's equation of Electrostatics by considering thermodynamic properties of free charges (ions). This equation is further linearized in practical applications [59].

Discarding electromagnetic constraints by the Geometrical model (Sec. 4) makes it simple and flexible on the one hand, and deprives it from meaningful physical constraints needed for identification of a unique solution, on the other hand. The unknown shape and effective density of the hydration shell [60] may also be estimated in optimization-based reconstructions utilizing the correlation between electron density and electrostatic potential.

Also, if the electron density and electrostatic potential are known, one can calculate the right PH and temperature. This may be helpful for the case where a crystal structure has been solved, and one seeks to introduce a chemical perturbation in the molecule. Knowing the perturbations in electron density and electrostatic potential, the required perturbation in the solvent properties can be estimated (in principle).

*7.5. Parameterized solutions and uniqueness of inverse X-ray scattering problems*

7.5.1. "Resolution", phase retrieval, and truncated scattering patterns

Even if single particle imaging can use the optical model and the "density map" used in protein crystallography, phase retrieval introduces a new challenge. Closely-related to phase retrieval is the notion of the maximum scattering angle (real-space resolution):
- In image processing, there exists the meaningful implicit assumption that the maximum spatial frequency of *data* is the maximum spatial frequency of the *object* generating data. In simple terms, the available spectrum has not been truncated (finite width and smooth terminations in both real- and Fourier-spaces).
- In protein crystallography, the retrieved real-space pattern is fitted to a model made up of known residues, and this model sets an effective absolute maximum frequency.
- In single particle imaging with truncated scattering patterns, the maximum spatial frequency in data $f_{Max}^{Data}$ corresponds to many different spatial frequencies of the object (including those higher than $f_{Max}^{Data}$). It simply originates from the nonlinearity of loss of phase. A radial component of the measured scattered intensity is related to different (including higher-order) radial components of the object; see Equation 15 in [61].

The crucial notion of "resolution" in single particle imaging has an ambiguity in whether it refers to data or the object. The consistent formulation in [61] or in the more critical evaluation of [62] assumes that the object itself has a maximum spatial frequency (corresponding to the order of basis functions). In many practical scenarios, however, the object has much higher frequencies compared to measured scattering data, and the measured scattering pattern is truncated. The existence of signatures from higher-order components of the object in measured low-order spatial frequencies makes the uniqueness of solution questionable.

This critical distinction between object/data resolution along with real/complex nature of the density map (and the meaning of density map in the first place) are all important physical aspects of single particle imaging that require (at least qualitative) consideration. Given the *established* status of phase retrieval in data processing, these issues (and possibly others) can be easily concealed. Despite its usefulness, *phase retrieval can be a serious source of generating spurious information and/or concealing the lack of a unique solution*. This fundamental issue is completely different from numerical challenges (non-convex optimization and being trapped in local minima, for which one may seek evolutionary or other smart algorithms).

7.5.2. *Unique* vs. *General* solution of inverse X-ray scattering

The uniqueness of the solution to an inverse X-ray scattering problem (or lack thereof) is a fundamental issue, which cannot be skipped by relying on smarter phase retrieval algorithms. *General* solutions can bring more reliability by sacrificing some conclusiveness. In comparison, in the theories of ordinary or partial differential equations, a *general* (parametric) solution is a meaningful answer to a problem without being unique. Further initial conditions or boundary conditions restrict the values of parameters to a smaller subset, and potentially to a unique one. In the established case of protein crystallography, systematic ways for enforcing such pieces of information is also a crucial part of data processing. Protein crystallography uses nontrivial optical scattering data to complete a puzzle (Section 5.3 in *Part 1*).

Whether single particle imaging is or is not able to find a unique solution independently, its reconstruction pipeline can benefit from parametric models to impose *transparent relevant controlled* type/level of *a-priori* information. Such formulations can also feature a modular design with disentangled contributions from *ab-initio* statistical analyses and *a-priori* geometrical/optical/chemical information, as in protein crystallography (Section 5.3 in *Part*

*1*). Analytical formulation of geometrical constraints, for instance in the case of icosahedral viral capsids, can reduce the number of parameters considerably in such a formulation.

*7.6. From protein crystallography to single-particle imaging*

The materials presented here in *Part 2* and also those in *Part 1* of this contribution can be summarized, as tabulated below in Table 2.

Table 2: The nontrivial journey from *Protein Crystallography* to *Single Particle Imaging* in a nutshell

|  | **Protein crystallography** | **Single particle imaging** | **Reference** |
|---|---|---|---|
| **Scattering pattern** | Discrete | Continuous |  |
| **Sample handling (delivery)** | Single secured rotating crystal | Random orientation, size, position |  |
| **Illumination gradients** | Robust | Vulnerable | Part 1, App. B |
| **Optical *a-priori* info** | Significant | Perceived unnecessary | Part 1, Sec. 5.3 |
| **Non-optical *a-priori* info** | Significant | Perceived unnecessary | Part 1, Sec. 5.3 |
| **Self-contained imaging** | No | Perceived possible | Part 1, Sec. 5.3 |
| **Scientific jurisdiction** | Chemistry | Optics | Part 1, Sec. 5.3 |
| **Modal illumination** | Yes | No | Part 1 App. C & Part 2, Sec. 4.2 |
| **Surface modes** | Negligible | Vulnerable | Part 2, Sec. 2.3, 7.3 |
| **Non-classical scattering** | No problem | Nontrivial signatures | Part 2, Sec. 5 |
| **Quantum confinement** | Nearly irrelevant | Relevant (core-shell …) | Part 2, Sec. 5.2, 6 |
| **Meaning/Type of "density map"** | Fixed, real | Diverse, complex | Part 2, Sec. 6 |
| **Phase retrieval ($q_{Max} = \infty$)** | Highly overdetermined, 2 measurements | Overdetermined, 1 measurement | Part 2, Sec. 6.6, 7.5 |
| **Highest resolution ($q_{Max}$)** | "Absolute" (fit to residues) | Coupled to missing orders | Part 2, Sec. 7.5 |

## 8. Conclusions

The efforts seeking to push the boundaries of X-ray *protein* crystallography towards single-particle imaging have already developed considerable expertise and achievements in the experimental and statistical parts. Theoretical considerations and hence the design/optimization/interpretation of experiments, however, have been mostly focused on the topic of "damage" and on *scientific* applications. Given the diversity of samples and illumination scenarios, there is a need for more rigorous treatment of *light propagation* in less-explored regimes of X-ray optics.

Compared to light-scattering experiments at optical (VIS/NIR) wavelengths, single-particle imaging with X-rays has a more ambitious goal (volumetric imaging rather than cross section calculation), yet a more primitive model of light propagation. This inconsistency seems to have been partially capitalized on the oversimplified outlook of protein crystallography. Furthermore, imaging big objects with soft X-ray photons makes the simple model of protein crystallography even less reliable.

Explicit differentiation between *volumetric* and *morphological* studies is expected to result in even more educated designs of experiments, more clarity in data analysis, and easier validation of and expectations from 3D reconstructions. More pronounced signatures of morphology in scattering patterns (compared to those of inclusions) can be further enhanced with surface waves and pulse excitations.

While less statistical bias is desirable, implicit assumptions (validity of the model of scattering, boundaries of the real-space support …) are meaningful, yet highly-biasing components of any statistical analysis. *Explicit* and *systematic* use of different types/levels of *a-priori* information (as done extensively in conventional protein crystallography) can narrow down the search space of 3D reconstructions in a meaningful process of elimination, even if *the* answer is not uniquely defined.

Statistical analyses have a crucial role in handling, reducing, and finding patterns in huge datasets. However, interpretation and specific quantitative use of statistical results for image reconstruction depend on specific theoretical model of scattering. Modeling and data analyses are complementary and need to *cross-validate* the *relevance* of each other. Rigorous validation schemes are required to avoid spurious (and possibly sought) features caused by untransparent statistical parameters and numerical errors. Promising results of statistical analyses in the context of single particle imaging are also helpful and motivating in taking over the complementary task (developing, simplifying, and assessing models), rather than keeping the ball of "theory" in the field of statistics.

Pessimistic interpretation of the issues addressed here as inevitable show-stoppers for single particle imaging is as far from objectivity as an optimistic view relying on the oversimplified outlook of protein crystallography. Some systematic studies may simply rule out or somehow handle many such issues for specific classes of objects, specific illumination scenarios, specific sample delivery schemes and so on. Appropriate level of surface roughness ($\lambda < \sigma_{roughness} < d_{resolution}$) may have a crucial role (similar to that of mosaicity for protein crystals) for "ideally-imperfect single particles".

**Appendices**

*Appendix A: Explicit formulation of scattering based on morphology*

In the framework of the common Geometrical model, the contribution of the density to the scattered electric field is written as a volume integral of the density function with a Fourier kernel. Assuming a piecewise homogeneous medium consisting of a big *shell* and smaller isolated embedded *cores*, one can write

$$F(\boldsymbol{q}) = \oiiint_{All} f(\boldsymbol{r})e^{-i2\pi \boldsymbol{q}\cdot\boldsymbol{r}}\, dV_r = \oiiint_{All} \rho_{Shell} e^{-i2\pi \boldsymbol{q}\cdot\boldsymbol{r}}\, dV_r + \sum_{Cores} \oiiint_{Core} (\rho_{Core} - \rho_{Shell})e^{-i2\pi \boldsymbol{q}\cdot\boldsymbol{r}}\, dV_r$$

The circles on triple integrals denote integration on a closed volume (simply-connected topology). For an arbitrary unit vector $\hat{a}$ (for instance for $\hat{a} = \hat{z}$ along the direction of X-ray propagation), we have

$$\oiiint e^{-i2\pi \boldsymbol{q}\cdot\boldsymbol{r}}\, dV_r = \frac{1}{-i2\pi \boldsymbol{q}\cdot\hat{a}} \oiint e^{-i2\pi \boldsymbol{q}\cdot\boldsymbol{r}}(\hat{a}\cdot d\boldsymbol{S}_r) = \frac{1}{-i2\pi q_z} \oiint e^{-i2\pi \boldsymbol{q}\cdot\boldsymbol{r}}(\hat{z}\cdot d\boldsymbol{S}_r)$$

So, the Fourier transform *for a locally homogeneous medium* can be written in terms of mere morphological information as

$$F(\boldsymbol{q}) = \rho_{Shell} \mathbb{M}_{Shell}(\boldsymbol{q}) + \sum_{Cores} (\rho_{Core} - \rho_{Shell})\, \mathbb{M}_{Core}(\boldsymbol{q})$$

where $\mathbb{M}_S(\boldsymbol{q}) = \frac{1}{-i2\pi q_z} \oiint_S e^{-i2\pi \boldsymbol{q}\cdot\boldsymbol{r}}(\hat{z}\cdot d\boldsymbol{S}_r)$ is a purely-morphological signature of the region enclosed by the surface $S$. While the shell signature $\mathbb{M}_{Shell}$ is scaled with the optical density $\rho_{Shell}$, the core densities $\mathbb{M}_{Core}$ are scaled with typically smaller values of $(\rho_{Core} - \rho_{Shell})$. Coherent inter-particle interference in $|F(\mathbf{q})|^2$ for a pair of objects with one *embedded* in the other (proportional to $\rho_1|\rho_1 - \rho_2|$) is weaker than the interference between the same objects when *isolated* (proportional to $\rho_1\rho_2$).

In first-order approximation, a small core contributes an offset proportional to its volume $\mathbb{M}_{Core}(\mathbf{q}) \sim V_{Core}$ to the scattered electric field, as

$$\oiint e^{-i2\pi\mathbf{q}\cdot\mathbf{r}}(\hat{z}.d\mathbf{S}_r) \sim \oiint [1 - i2\pi\mathbf{q}.\mathbf{r}](\hat{z}.d\mathbf{S}_r) = \oiiint \nabla.[(1 - i2\pi\mathbf{q}.\mathbf{r})\hat{z}]dV_r = -i2\pi q_z V$$

Small sizes and differential densities are two factors that suppress the signatures of inclusions in the scattered field, compared to that of the shell.

Fitting a scattering pattern to a real-space parametric model can benefit from 2D (morphological) formulation of the full 3D scattering pattern $\mathbb{M}_{Shell}(\mathbf{q})$. It reduces the number of the sought parameters from $O(N^3)$ to $O(N^2)$, where $N$ is the number of grid points per coordinate (or basis functions per parameter). Specifically, it makes a formulation based on spherical harmonics (per shell or core) sufficient; with no need to radial basis functions.

Consider a convex homogeneous solid with a unity density and *embedded in the unit sphere*, modeled as $f(\mathbf{r}) = u[r_{surf}(\Omega_r) - r]$, where $u(x)$ is the Heaviside step function. Morphological information $r_{surf}(\Omega_r)$ as a function of real-space spherical coordinate angles $\Omega_r = (\theta_r, \phi_r)$ is coupled to similar angular variations in the Fourier space; forming a Maclaurin series in terms of $q = |\mathbf{q}|$:

$$F(\mathbf{q}) = \oiint d\Omega_r \int_{r=0}^{r=r_{surf}(\Omega_r)} r^2 e^{-i2\pi qr\cos(\gamma)} dr = \sum_{n=0}^{\infty} \frac{(-i2\pi q)^n}{(n+3)n!} F_n(\Omega_q)$$

$$F_n(\Omega_q) = \oiint d\Omega_r \cos^n(\gamma(\Omega_r, \Omega_q)) r_{surf}^{n+3}(\Omega_r)$$

Where $\cos(\gamma) = \cos(\theta)\cos(\theta_q) + \sin(\theta)\sin(\theta_q)\cos(\phi - \phi_q)\gamma$. For the unit sphere $r_{surf}(\Omega_r) = 1$, $F_n(\Omega_q) = [1 + (-1)^n]/(n+1)$ independent of $\Omega_q$, and the above series can be rewritten in closed-form as the known spherical scattering $F(\mathbf{q}) = [sin(2\pi q) - (2\pi q)cos(2\pi q)]/(2\pi q)^3$. This formulation quantifies the way 3D Fourier-space profile is correlated to mere 2D (morphological) profile in real-space.

*Appendix B: Correction for accumulated extinction*

With small scattering angles, assuming the validity of the Geometrical model, and using the Fourier-Slice theorem, the 2D scattered field is related to the 2D Fourier transform of the projected density along the X-ray. The density is twice the index contrast $\rho = 2(1 - n)$, and hence the real-space projection and the scattered intensity can be written as

$$E_p(x,y) = \int_{z=-\infty}^{z=+\infty} 2[1 - n(x,y,z)]dz \quad \& \quad I(q_x, q_y) \propto \left| \iint e^{-i2\pi\vec{r}_\|\cdot\vec{q}_\|} E_p(x,y) dxdy \right|^2$$

With weak interactions in infinitesimally-thin longitudinal slices, there will be a plane-wave-like propagation $E(z + dz) = E(z)e^{-i2\pi k_0 ndz}$, or $E(z) = E_{-\infty}e^{-i2\pi k_0 \int_{z'=-\infty}^{z'=z} n(x,y,z')dz'}$. By changing the density to illuminated density, discarding the linear phase shift of propagation (absorbed in the Fourier kernel), and defining $L(x,y,z) \equiv \int_{z'=-\infty}^{z'=z}[1 - n(x,y,z')]dz'$, the projected density and its first-order approximation can be written as

$$E_p(x,y) = \int_{z=-\infty}^{z=+\infty} e^{-i2\pi k_0 L(x,y,z)} 2[1 - n(x,y,z)]dz$$

$$E_p(x,y) = 2\int_{z=-\infty}^{z=+\infty}[1-i2\pi k_0 L]L'dz = 2[(1-i2\pi k_0 L)L]_{-\infty}^{+\infty} - 2\int_{z=-\infty}^{z=+\infty}L[-i2\pi k_0 L']dz$$

$$= 2(1-i2\pi k_0 L_{+\infty})L_{+\infty} - 2\int_{z=-\infty}^{z=+\infty}[1-i2\pi k_0 L]L'dz + 2\int_{z=-\infty}^{z=+\infty}L'dz$$

$$= 2(1-i2\pi k_0 L_{+\infty})L_{+\infty} - E_p(x,y) + 2L_{+\infty}$$

But $2L_{+\infty}$ is simply the ideal projection unaffected by extinctions. So:

$$\boldsymbol{E_p(x,y) = (1 - i\pi k_0 E_{ideal})E_{ideal} = \left(1 - \frac{i\pi}{\lambda}E_{ideal}\right)E_{ideal}}$$

Ideally, $E_p(x,y)$ is simply the accumulated (projected) extinction or $E_{ideal} = 2L_\infty$. When not, the above expression provides a first-order correction scheme. The quadratic equation can also be solved to retrieve the ideal projection from experimental projection:

$$E_{ideal} - \frac{i\pi}{\lambda}E_{ideal}^2 - E_p = 0 \rightarrow E_{ideal}^2 + \frac{i\lambda}{\pi}E_{ideal} + -\frac{i\lambda}{\pi}E_p = 0$$

$$E_{ideal} = \left(\frac{i\lambda}{2\pi}\right)\left[-1 \pm \sqrt{1 + E_p/(\frac{i\lambda}{\pi})}\right] \sim \left(\frac{i\lambda}{2\pi}\right)\left[-1 \pm \left(1 + E_p/(\frac{i\lambda}{2\pi}) - 0.25E_p^2/\left(\frac{i\lambda}{2\pi}\right)^2\right)\right]$$

It is easy to choose the right sign (positive), and one will have

$$\boldsymbol{E_{ideal} = E_p - \frac{0.25}{\frac{i\lambda}{2\pi}}E_p^2 = \left(1 + \frac{i\pi}{2\lambda}E_p\right)E_p}$$

*Appendix C: Disentangling the contributions of core and shell in scattering patterns*

Let $v(x) = 1$ for $0 \leq x \leq 1$ and 0 otherwise. Also let the core and shell densities be limited to the sphere $0 \leq r \leq R_1$ and the spherical shell $R_1 \leq r \leq R_2$, respectively and have densities $f^c(r)$ and $f^s(r)$ defined over the entire $0 \leq r \leq R_2$ range. Expansion of 3D functions and their Fourier transforms using spherical harmonics results in

$$f(\boldsymbol{r}) = f^s(\boldsymbol{r})[v(r/R_2) - v(r/R_1)] + f^c(\boldsymbol{r})v(r/R_1)$$

$$f(\boldsymbol{r}) = \sum_{l,m}\left\{f_{l,m}^c(r)v\left(\frac{r}{R_1}\right) + f_{l,m}^s(r)\left[v\left(\frac{r}{R_2}\right) - v\left(\frac{r}{R_1}\right)\right]\right\}Y_{l,m}(\Omega_r)$$

We also assume a spherical harmonic expansion of the phased 2D snapshot (on the Ewald sphere), as

$$\tilde{F}(\boldsymbol{q}) = \sum_{l,m}\tilde{F}_{l,m}(q)Y_{l,m}(\Omega_q)$$

We use an orthonormal basis $\phi_s(r)$ to describe the real-space profiles of $f_{l,m}^c(r)$ and $f_{l,m}^s(r)$. We also define a function $\eta$ to correlate the *known* real-space functions $\phi_s(r)$ to their reciprocal-space counterparts. Finally, we define two mutually-orthogonal bases $\psi$ and $\chi$ (say sine/cosine) to describe $\tilde{F}_{l,m}(q)$. Given the orthogonality of spherical harmonics $Y_{l,m}(\Omega_q)$, we need to match the coefficients of $\tilde{F}(\boldsymbol{q})$ and $F(\boldsymbol{q})$ for different $(l,m)$ pairs, separately. As such, we drop the trivial indices $l,m$ for the coefficients of the following expansions and perform 3D Fourier transforms [62] as follows:

$$F(\boldsymbol{q}) = 4\pi\sum_{l,m}(-i)^l\left[\int_0^{R_1}f_{l,m}^c(r)j_l(2\pi qr)r^2dr + \int_{R_1}^{R_2}f_{l,m}^s(r)j_l(2\pi qr)r^2dr\right]Y_{l,m}(\Omega_q)$$

$$\eta_l^s(q,x) \equiv 4\pi(-i)^l\int_0^x\phi_s(r)j_l(2\pi qr)r^2dr$$

$$f_{l,m}^s(r) = \sum_s p_s\phi_s(r)$$

$$f_{l,m}^c(r) = \sum_s q_s\phi_s(r)$$

$$\tilde{F}_{l,m}(q) = \sum_n[C_n\psi_n(q) + D_n\chi_n(q)]$$

Matching the coefficients of the phased measured snapshot and the estimated scattering results in

$$\sum_n [C_n \psi_n(q) + D_n \chi_n(q)] = \sum_s [p_s \eta_i^s(q, R_1) + q_s(\eta_i^s(q, R_2) - \eta_i^s(q, R_1))]$$

We denote the weight functions associated with the orthonormal basis functions $\psi/\chi$ by $w_\psi/w_\chi$. Multiplying both sides once by $\psi_m^*(q)w_\psi(q)$ and once by $\chi_m^*(q)w_\chi(q)$ and using the linearity and orthonormality of the basis functions, we have

$$C_m = \sum_s [p_s \langle \psi_m(q) | w_\psi | \eta_i^s(q, R_1) \rangle + q_s \langle \psi_m(q) | w_\psi | \eta_i^s(q, R_2) - \eta_i^s(q, R_1) \rangle]$$

$$D_m = \sum_s [p_s \langle \chi_m(q) | w_\chi | \eta_i^s(q, R_1) \rangle + q_s \langle \chi_m(q) | w_\chi | \eta_i^s(q, R_2) - \eta_i^s(q, R_1) \rangle]$$

The expressions $\langle f | w | g \rangle$ are simply the *constant* projections of the *known* basis functions on each other. They depend on the reconstruction parameters $R_1, R_2, q_{max}$, but independent from measured data. Also note that changing the number of the employed basis functions (increasing the resolution) does not change the $\langle f | w | g \rangle$ numbers, as they are derived from orthogonal projections (as opposed to nonlinear order-dependent fits).

By putting similar coefficients $p_s, q_s, C_m, D_m$ together as vectors $\boldsymbol{P}, \boldsymbol{Q}, \boldsymbol{C}, \boldsymbol{D}$, the above equations are simply written as a linear matrix equation in terms of the vectors

$$\begin{bmatrix} \alpha & \beta \\ \delta & \omega \end{bmatrix} \begin{bmatrix} \boldsymbol{P} \\ \boldsymbol{Q} \end{bmatrix} = \begin{bmatrix} \boldsymbol{C} \\ \boldsymbol{D} \end{bmatrix}$$